\documentclass[letterpaper,12pt]{article}

\usepackage{booktabs}
\usepackage{multirow}
\usepackage{pdflscape}
\usepackage{rotating}
\usepackage{xcolor}
\usepackage{fancyvrb}

\usepackage{graphicx}

\input{preambleKnit.tex}

\usepackage[authoryear,comma,longnamesfirst,sectionbib]{natbib}
\usepackage{amsmath}%
\usepackage{amsfonts}%
\usepackage{amssymb}%
\usepackage{booktabs}
\usepackage{url}
\usepackage{listings}
\usepackage{breakurl}
\newcommand{\bse}{\textbf{\textrm{b}}}

\newcommand{\bsb}{\textbf{\textrm{b}}}
\newcommand{\bsbb}{\textrm{b}}
\newcommand{\bsaa}{\textrm{a}}
\newcommand{\bsZ}{\boldsymbol{Z}}
\newcommand{\bsW}{\boldsymbol{W}}
\newcommand{\bsS}{\boldsymbol{S}}
\newcommand{\bsalpha}{\boldsymbol{\alpha}}
\newcommand{\bsbeta}{\boldsymbol{\beta}}

\newcommand{\bsy}{\boldsymbol{y}}
\newcommand{\bsr}{\boldsymbol{r}}
\newcommand{\bsR}{\boldsymbol{R}}
\newcommand{\bsX}{\boldsymbol{X}}
\newcommand{\bsG}{\boldsymbol{G}}
\newcommand{\ud}{\mathrm{d}}
\newcommand{\bds}[1]{\boldsymbol{#1}}
\newcommand{\te}[1]{\textrm{#1}}

\title{Multivariate Generalized Linear Mixed Models for Joint Estimation of Sporting Outcomes}
\date{}
\author{Jennifer E. Broatch\thanks{Corresponding Author, Arizona State University} \and  Andrew T. Karl \thanks{Adsurgo LLC} }

\begin{document}

\maketitle

\begin{abstract}
This paper explores improvements in prediction accuracy and inference capability when allowing for potential correlation in team-level random effects across multiple game-level responses from different assumed distributions. First-order and fully exponential Laplace approximations are used to fit normal-binary and Poisson-binary multivariate generalized linear mixed models with non-nested random effects structures. We have built these models into the R package mvglmmRank, which is used to explore several seasons of American college football and basketball data.
\end{abstract}

\section*{NOTICE}
This is the author's version of a work that was accepted for publication in the \textit{Italian Journal of Applied Statistics}. Changes resulting from the publishing process, such as peer review, editing, corrections, structural formatting, and other quality control mechanisms may not be reflected in this document. Changes may have been made to this work since it was submitted for publication.

\section{Introduction} \label{sec:intro}
 Traditionally, algorithms for ranking sports teams and predicting sporting outcomes utilize either the observed margin of victory (MOV) \citep{henderson75} or the binary win/loss information \citep{mease,football}, along with potential covariates such as game location (home, away, neutral).
 In contrast, we jointly model either MOV or win/loss along with a separate game-level response, which is shown to improve predictions under certain model specifications. We present a set of non-nested generalized linear mixed models to jointly model the MOV or win/loss along with a game outcome, such as penalty yards or number of penalties, shots on goal, turnover margin.
 Multiple response distributions are necessary to model a variety of sporting outcomes and are available in the model and presented R package for executing the model. For example, the normal distribution is most suitable to model the score in a high scoring sport such as basketball, where as a Poisson model may be more appropriate for scores in hockey or soccer (football).

  In this paper,  we explore the benefits of modeling these responses jointly, assuming conditional independence given correlation between distinct team effects for each response. For some responses, the joint models benefit from significantly improved median log-loss and absolute residuals of cross-validation predictions. Furthermore, the joint model provides the ability to test for significant relationships between high-level hierarchical effects (e.g. random team effects) since significant predictors for outcomes at the game level may not be important at the team level. We have published our R \citep{R} code for these models on CRAN (\url{http://cran.r-project.org/}) via the package mvglmmRank: the appendix provides a demonstration of the package. The  data used to produce the results in this paper are made available at \burl{github.com/10-01/NCAA-Football-Analytics}.

  Previous works have also considered the joint modeling of team ratings and outcome prediction.
  \citet{annis} present a two-stage, hierarchical ``hybrid ranking system'' that can be considered an average of win/loss and point-scoring models, focusing on the prediction of NCAA football rankings.
  In stage 1, the win/loss indicator is modeled. In stage 2, the scores are predicted conditioned on the win/loss outcome in stage 1. Each team is modeled with an offensive (fixed) effect and a defensive (fixed) effect. Model estimation relies on generalized estimating equations (GEE).  The win/loss indicators are modeled by comparing the ``merit'' of each team, which is defined as the sum of the offensive and defensive ratings. \citet{baio} use a similar point-scoring model (in a Bayesian framework), fitting separate ``attack'' and ``defense'' values for each team.
Several other papers have considered modeling separate offense and defense effects \citep{karlis,baio,ruiz}.

While \citet{annis} use the sum of a team's offensive and defensive effects to represent their winning propensity in a logistic regression, we build upon the Poisson-binary model proposed by \citet{karlcgs} and fit a separate win-propensity random effect for each team. This effect is correlated with, rather than determined by, the offensive and defensive effects from the point-scoring (yards-recorded, etc) model. These three team-level effects are modeled as random effects in a multivariate generalized linear mixed model (GLMM). This allows us to measure and compare the relationships between offensive/defensive ability (with respect to a variety of responses) and winning propensity. The binary win/loss indicators are jointly modeled with the team scores or other responses by allowing the random team effects in the models for each response to be correlated. Assuming a normal distribution for the team effects imposes a form of regularization, allowing the binary model to be fit in the presence of undefeated or winless teams \citep{football}.

Suppose a binary win/loss indicator is jointly modeled, then for the binary win/loss indicators, an underlying latent trait or ``win-propensity'' rating is assigned to each team.
These ratings are fit simultaneously with two game level  response-propensity ratings: offensive and defensive.
The model presented allows for potential correlation between all three ratings by fitting them using random effects assuming a multivariate normal distribution.
To illustrate, this paper examines how the joint modeling of win/loss indicators and four different game-level responses in American football (yards per play, sacks, fumbles, and score described in section \ref{sec:joint})
may lead to an improvement in the cross validation predictions of both responses versus the traditional model.
The results indicate that a higher correlation between the win/loss response and the game-level response lead to improved cross-validated predictions.

Section~\ref{sec:model} describes a set of multivariate generalized linear mixed models for predicting game outcomes, and section~\ref{sec:computation} describes the computational approach used by the mvglmmRank package. Section~\ref{sec:ncaaf} compares cross-validation prediction accuracy of several game-level responses across three American college football seasons. Section~\ref{sec:ncaapred} evaluates the performance of the joint model across nineteen college basketball (NCAA) tournaments.  Appendix~\ref{sec:bb} provides a demonstration of the implementation of the joint model in the mvglmmRank R package.

\section{The Model}\label{sec:model}

When modeling $n$ games, let $r_i$ be a binary indicator for the outcome of the $i$-th game for $i=1,\ldots,n$, taking the value 1 with a home team ``win'' and 0 with a visiting team ``win'', where ``win'' can be defined to be outscoring the opponent, receiving fewer penalties than the opponent, etc. A neutral-site indicator  is used to indicate that the home team was designated arbitrarily. The home-win indicators are concatenated into the vector $\bsr=(r_1,\ldots,r_n)^{\prime}$. We will use $y_{ih}$ to denote the score (or penalties, yards-per-play, etc.) of the home team in the $i$-th game, and $y_{ia}$ to denote the score of the away team, letting $\bsy_i=(y_{ih},y_{ia})^{\prime}$. The scores are concatenated into the vector $\bsy=(\bsy_1^{\prime},\ldots,\bsy_n^{\prime})^{\prime}$. We will assume separate parametric models for $\bsy$ and $\bsr$; however, these models will be related by allowing correlation between the random team effects present in each model.

Suppose we wish to model the outcome of a game between the home team, H, and the away team, A. We assume that each team may be described by three potentially related characteristics: their offensive rating ($\bsbb^o$), defensive rating ($\bsbb^d$), and a rating ($\bsbb^w$) that quantifies their winning propensity. Heuristically, we want to find the ratings that satisfy
\begin{align*}
\te{E}[y_{ih}]=&f_1(\bsbb^o_h-\bsbb^d_a)\\
\te{E}[y_{ia}]=&f_1(\bsbb^o_a-\bsbb^d_h)\\
P(r_i=1)=&f_2\left(\bsbb^w_h-\bsbb^w_a\right)
\end{align*}
for some functions $f_1$ and $f_2$. To do this, we will specify the functions $f_1$ and $f_2$, the assumed distribution of $\bsy$ conditional on the offensive and defensive ratings, the assumed distribution of $\bsr$ conditional on the win propensity ratings, and the assumed distribution of (and relationship between) the ratings. Due to the binary nature of $r_i$, $f_2$ will necessarily be a nonlinear function. The offense and defense ratings for each team are calculated while controlling for the quality of opponent, implicitly considering strength of schedule as in \citet{harville77}, \citet{annis}, and \citet{football}. By contrast, raw offensive and defensive totals inflate the ranking of teams that play a set of easy opponents and penalize those that play a difficult schedule.

We model the offensive, defensive, and win propensity ratings of the $j$-th team for $j=1,\ldots,p$ with random effects $\bsbb_j^o$, $\bsbb_j^d$, and $\bsbb_j^w$ assuming $\bsb_j=(\bsbb_j^o,\bsbb_j^d,\bsbb_j^w)^{\prime}\sim N_3(\bds{0},\bsG^*)$, where $\bsG^*$ is an unstructured covariance matrix and $p$ represents the number of teams being ranked.  In addition, $\bsb\sim N(\bds{0},\bsG)$, where $\bsb=(\bsb_1^{\prime},\ldots,\bsb_p^{\prime})^{\prime}$ and $\bsG$ is block diagonal with $p$ copies of $\bsG^*$. We allow $\bsy|\bse$ to follow either a normal or a Poisson distribution. While we use $\bsy|\bse$ as a notational convenience, we do not condition $\bsy$ on $\bse^w$. Likewise, we will use $\bsr|\bse$ when we may more explicitly write $\bsr|\bse^w$.

\subsection{Bivariate Normal Outcomes}\label{sec:bivariate}
We may assume a bivariate normal distribution for the outcomes (e.g. scores) of the $i$-th game $\bsy_i|\bse\sim N_2(\bsX_i\bsbeta+\bsZ_i\bsb,\bsR^*)$. In the error covariance matrix, $\bsR^*$, we model the potential intra-game correlation between the responses of opposing teams: the (1,1) term models the conditional variance of the home team responses, the (2,2) term models the conditional variance of the away team responses, and the (1,2)=(2,1) term models the conditional covariance of the home and away team responses. $\bsy|\bse\sim N_{2n}(\bsX\bsbeta+\bsZ\bsb,\bsR)$, where $\bsR$ is block diagonal with $n$ copies of $\bsR^*$, and $\bsX$ and $\bsZ$ are the concatenation of the $\bsX_i$ and $\bsZ_i$, which are defined below. $\bsbeta$ may be used to model any fixed effect covariates, though we only consider a parsimonious model with a mean and a home field effect, that is, $\bsbeta=(\beta_h,\beta_a,\beta_n)^{\prime}$ where $\beta_h$ is the mean home response, $\beta_a$ is the mean away response, and $\beta_n$ is the mean neutral site response. The design matrix $\bsX_i$ is a $2\times 3 $ matrix with an indicator for the ``home'' team in the first row and for the ``away'' team in the second row. If the home and away teams were designated arbitrarily for a neutral site game, then \[\bsX_i=\left(\begin{array}{ccc} 0&0&1\\0&0&1 \end{array}\right).\] The error terms of the arbitrarily designated teams are still modeled with the corresponding ``home'' and ``away'' components of $\bsR^*$, but the relative infrequency of neutral site games in most applications minimizes any impact this may have. Even if every game in the data set is a neutral site game, $\widehat{\bsR^*}$ will still be unbiased (since the selection of the ``home'' team is randomized), though inefficient (since two parameters are being used to estimate the same quantity, halving the sample size used to estimate each parameter). In such situations, the two diagonal components of $\bsR^*$ should be constrained to be equal.

$\bsZ_i$ is a $2 \times 3p$ matrix that indicates which teams competed in game $i$. If team $k$ visits team $l$ in game $i$, then in its first row, $\bsZ_i$ contains a 1 in the position corresponding to the position of the offensive effect of team $l$, $\bsbb^o_l$, in $\bsb$, and a $-1$ in the position corresponding to the position of the defensive effect, $\bsbb^d_k$, of team $k$. In its second row, $\bsZ_i$ contains a 1 in the position corresponding to the position of the offensive effect of team $k$, $\bsbb^o_k$, in $\bsb$, and a $-1$ in the position corresponding to the position of the defensive effect, $\bsbb^d_l$, of team $l$. This is a multiple membership design \citep{browne01} since each game belongs to multiple levels of the same random effect. As a result, $\bsZ$ does not have a patterned structure and may not be factored for more efficient optimization, as it could be with nested designs. The likelihood function for the scores under the normally distributed model is

\footnotesize
\begin{align}\label{eq:normal}
f(\bsy|\bse)=&\prod_{i=1}^n  \left[(2\pi)^{-1}|\bsR^*|^{-1/2}\te{exp}\left\{-\frac{1}{2}(\bsy_{i}-\bsX_{i}\bsbeta+\bsZ_{i}\bse)^{\prime}{\bsR^*}^{-1}(\bsy_{i}-\bsX_{i}\bsbeta+\bsZ_{i}\bse)\right\}\right].
\end{align}
\normalsize
This is a generalization of the mixed model proposed by \citep{harville77} for rating American football teams.

\subsection{Two Poisson Outcomes}
We may alternatively assume a Poisson distribution for the conditional responses (e.g. scores, turnovers). When modeling $\bsy|\bse$ using a GLMM with a Poisson distribution and the canonical log link, it is not possible to model the intra-game correlation with an error covariance matrix since the variance of a Poisson distribution is determined by its mean. Instead, we may optionally add an additional game-level random effect, $\bsaa_i$, and thus an additional variance component, $\sigma^2_g$, to $\bsG$. In this case, we recast $\bsb$ as $\bsb=(\bsb_1,\ldots,\bsb_p,\bsaa_1,\ldots,\bsaa_n)^{\prime}$ and $\bsG=\te{block diag}(\bsG^*,\ldots,\bsG^*,\sigma^2_g I_n)$.

\begin{align*}
y_{i*}|\bse&\sim \te{Poisson}(\mu_{i*})\\
\log(\mu_{i*})&=\bsX_{i*}\bsbeta+\bsZ_{i*}\bse
\end{align*}
where $*$ may be replaced by $h$ or $a$. Regardless of whether or not the game-level effect is included, the likelihood function may be written as
\begin{equation}\label{eq:Poisson}
f(\bsy|\bse)=\prod_{i=1}^n \prod_{*\in\{a,h\}} \left[\frac{1}{y_{i*}!}\te{ exp}\left\{y_{i*}(\bsX_{i*}\bsbeta+\bsZ_{i*}\bse)\right\}\te{exp}\left\{-\te{exp}\left[\bsX_{i*}\bsbeta+\bsZ_{i*}\bse\right]\right\}\right].
\end{equation}
{ For high-scoring sports such as basketball, the Poisson distribution is well approximated by the normal. However, the option to fit Poisson scores will remain useful when modeling low-scoring sports (e.g. soccer, baseball, hockey) or low-count outcomes such as number of penalties, as discussed in section~\ref{sec:ncaaf}}.

\subsection{Binary Outcomes}
Rather than modeling the team scores resulting from each contest, we may model the binary win/loss indicator for the ``home'' team. Predictions for future outcomes are presented as the probability of Team H defeating Team A, as opposed to the score predictions for each team that are available when modeling the scores directly.  \citet{football} considers multiple formulations of a multiple membership generalized linear mixed model for the binary outcome indicators: we will focus on one of those. Letting $\pi_i=P(r_i=1)$, we model the probability of a home win with a GLMM assuming a Bernoulli conditional distribution and use a probit link,
\begin{align*}\label{eq:binary}
r_i|\bse&\sim \te{Bin}(1,\pi_i)\\
\Phi^{-1}(\pi_i)&=W_i\alpha+\bsS_i\bse
\end{align*}
where $\Phi$ denotes the normal cumulative distribution function. Ties are handled by awarding a win (and thus a loss) to each team.

The home field effect is measured by $\alpha$, with a coefficient vector $\bsW$. $W_i$ takes the value 0 if the $i$-th game was played at a neutral site and 1 otherwise. The design matrix $\bsS$ for the random effects contains rows $\bsS_i$ that indicate which teams competed in game $i$. If team $k$ visits team $l$ in game $i$, then $\bsS_i$ is a vector of zeros with a $1$ in the component corresponding to the position of $\bsbb_l^w$ in $\bsb$ and a $-1$ in the component corresponding to $\bsbb_k^w$. Note that $\bsr$ is conditioned only on $\bsb^w$, and not on $(\bsb^o,\bsb^d)$. Pragmatically, all of the components in the columns of $\bsS$ corresponding to the positions of $\bsb^o$ and $\bsb^d$ in $\bsb$ are 0. The likelihood function for the binary indicators is
\begin{equation}\label{eq:binary}
f(\bsr|\bse)=\prod_{i=1}^n \left[\Phi\left\{\left(-1\right)^{1-r_i}\left[W_i\bsalpha+\bsS_i\bse\right]\right\}\right].
\end{equation}

\subsection{The Joint Model}\label{sec:joint}
Traditionally, teams ratings have been obtained by maximizing only one of the likelihoods (\ref{eq:normal}), (\ref{eq:Poisson}), or (\ref{eq:binary}). { \citet{karlcgs} propose the joint Poisson-binary model for team scores and game outcomes, focusing on the derivation of computational details, which are summarized in the next section. In this paper, we consider more general applications to other game-level responses.} The joint likelihood function \ref{eq:joint} 
\begin{align}\label{eq:joint}
L(\bsbeta,\bsG,\bsR)&=\idotsint f(\bsy|\bse) f(\bsr|\bse) f(\bse) \ud \bse
\end{align}
simultaneously maximizes (\ref{eq:binary}) along with a choice of either (\ref{eq:normal}) or (\ref{eq:Poisson})
where $f(\bse)$ is the density of $\bse\sim N(\bds{0},\bsG)$. 
The key feature of the joint model is the pair of off-diagonal covariance terms between $(\bsb^o,\bsb^d)^{\prime}$ and $\bsb^w$ in the $\bsG$ matrix. If these covariance terms were constrained to 0 then the resulting model fit would be equivalent to that obtained by modeling the two responses independently.
Thus, the joint model contains the individual normal/Poisson and binary models as a special case: the additional flexibility afforded by Model (\ref{eq:joint}) may lead to improved predictions for both responses when team win-propensities are correlated with their offensive and defensive capabilities. A similiar normal-binary correlated random effects model was employed by \citet{karlcpm} in order to jointly model student test scores in a value-added model with binary attendance-indicators in order to explore sensitivity to the assumption that data were missing at random.

 In addition to fitting each of the response types described in the previous subsections individually, the mvglmmRank package offers options to fit the joint normal-binary and Poisson-binary models.
Just as the individual score and outcome models may make opposite predictions about the game outcome, the joint model occasionally will predict a team to outscore its opponent in the score model while also predicting less than a 50\% chance of that team winning. This is a result of modeling distinct team rather than constraining them to be equal to the sum of offensive and defensive ratings, as done by \citet{annis}. The benefit of this approach is that the relative strength of the defense/win-propensity and offense/win-propensity correlations may be compared. The outcomes predicted by the binary component of the joint model focus on the observed win/loss outcomes while allowing the team win-propensity ratings to be influenced by the team offensive and defensive ratings. On the other hand, the outcomes predicted by the score component (checking which team has a larger predicted score) give a relatively larger weight to the observed scores, making the predictions susceptible to teams running up the score on weak opponents \citep{harville03}. As demonstrated in section~\ref{sec:ncaapred}, the joint model tends to produce improved probability estimates over those produced by the binary model.

\section{Computation}\label{sec:computation}
The likelihood functions in Equations (\ref{eq:Poisson}), (\ref{eq:binary}) and (\ref{eq:joint}) contain intractable integrals because the random effects enter the model through a nonlinear link function. Furthermore, the $p$-dimensional integral in each equation may not be factored as a product of one-dimensional integrals. Such a factorization occurs in longitudinal models involving nested random effects. However, the multiple membership random effects structure of our model results in a likelihood that may not be factored. It is possible to fit multiple membership models in SAS, using the EFFECT statement of PROC GLIMMIX. \citet{football} provides code for fitting the binary model in GLIMMIX. There are, however, advantages to using custom-written software instead. Building the model fitting routine into an R package makes the models available to readers who do not have access to SAS. 
Secondly, GLIMMIX does not currently account for the sparse structure of the random effects design matrices, resulting in exponentially higher memory and computational costs than are required when that structure is accounted for \citep{karlem}. Thirdly, the EM algorithm may be used to provide stable estimation in the presence of a near-singular $\bds{G}$ matrix \citep{karlem,karlcgs}, whereas GLIMMIX relies on a Newton-Raphson routine that tends to step outside of the parameter space in such situations.

Finally, we are able to use more accurate approximations than the default pseudo-likelihood approximation \citep{wolfinger93} of GLIMMIX, including first-order and fully exponential Laplace approximations \citep{tierney89,karlcgs}. (GLIMMIX is capable of using the first-order Laplace approximation, but we have not had success using it with the EFFECT statement). In line with the theory and simulations presented by \citet{karlcgs}, 17 of the 18 basketball tournaments in section~\ref{sec:ncaapred} are modeled more accurately in the binary model as fully exponential corrections are applied to the random effects vector. In those same seasons, the predictions show further improvement with the addition of fully exponential corrections to the random effects covariance matrix.

\citet{karlcgs} describe the estimation of multiple response generalized linear mixed models with non-nested random effects structures and derive the computational steps required to estimate the Poisson-binary model with an EM algorithm. The models presented here are special cases of that class of models. The exact maximum likelihood estimates are obtained for the normal model when team scores are modeled alone. The mvglmmRank package implements these methods without requiring end-user knowledge of the estimation routine.
Section~\ref{sec:bb} demonstrates the use of the package in the context of modeling  college football yards-per-game with home-win indicators.

The mvglmmRank package reports the Hessian of the parameter estimates. The inverse of this matrix is an estimate for the asymptotic covariance matrix of the parameter estimates, and it ought to be positive-definite \citep{demidenko}.  A singular Hessian suggests that the model is empirically underidentified with the current data set \citep{rabe01}. This can be caused by a solution on the boundary of the parameter space (e.g. zero variance components, linear dependence among the random effects), by multicollinearity among the fixed effects, convergence at a saddle point, or by too loose of a convergence criterion. \citet{rabe01} recommend checking the condition number (the square root of the largest to the smallest eigenvalue) of the Hessian. However, the Hessian is sensitive to the scaling of the responses, while the correlation matrix of the inverse Hessian (if it exists) is invariant. As such, we prefer to check the condition number of this correlation matrix. While the joint model for scores and win/loss outcomes for the data set presented by \citet{karlcgs} does not show signs of empirical underidentification, this model does show such signs for other data sets when modeling scores and win/loss outcomes. This seems reasonable, since the win/loss indicators are simply a discretized difference of the team scores. While the model parameters are unstable in the presence of empirical underidentification, the predictions produced by the model remain useful as evidenced by improvement in cross validation error rates, a point discussed in section~\ref{sec:ncaaf}.  Joint models for other responses (e.g. fumbles) with the win/loss indicators do not typically show symptoms of underidentification.

\section{American College Football Outcomes}\label{sec:ncaaf}
This section considers several different game-level outcomes from the 2005--2013 American College Football seasons. The models presented here are fit independently across each of the nine seasons.
The data were originally furnished under an open source license from \url{cfbstats.com}, and are now maintained at \url{https://github.com/10-01/NCAA-Football-Analytics}. As mentioned in section \ref{sec:intro}, the standard modeling outcomes margin of victory (MOV) \citep{henderson75} and the binary win/loss information \citep{mease,football}, along with potential covariates such as game location (home, away, neutral) will be used.
To illustrate the joint model, we will use recordings for game-level responses: sacks, yards per play, and fumbles in addition to the game-scores.
For those unfamiliar with American Football, a ``sack'' is recorded when a defensive player ``sacks'' the quarterback, who receives the ball to begin a play, before they are able to make a positive move forward toward the goal.
A ``sack'' has a positive impact on the defensive ability of a team.
Sacks are relatively infrequent.  The leading American Football teams average about 3 sacks per game.
``Yards per play'' is calculated by an offensive move towards the goal, regardless of type of play (e.g. run or pass).
A football field is 100 yards, and a team has 4 attempts to move the football 10 yards down the field at a time.
 A higher value of ``yards per play'' would indicate a higher offensive ability.
 A ``fumble'' is recorded when a player loses the ball on the ground and either team is able to pick it up.
 A ``fumble lost'' would indicate a turnover to the other team.
 This paper will use ``fumbles'' rather than ``fumbles lost'' to demonstrate the effectiveness of the joint model when a low correlated or irrelevant response is used.
We have made the processed data for each season available.

To compare the effectiveness of the various models, the predictions for the home-win indicator for game $i$ are scored against the actual game outcomes using a log-loss function: 
\begin{equation}
\te{log-loss}_{i}=-y_i\log\left(\hat{y}_i\right)-\left(1-y_i\right)\log\left(1-\hat{y}_i\right)
\end{equation}
where $\hat{y}_i$ is the predicted probability of a home-team win in game $i$, $y_i$ is the outcome of game $i$ (taking the value 1 with a home-team victory and 0 otherwise). A smaller value of log-loss represents a more accurate prediction.

Using 10-fold cross-validation for each of the seasons, we compare the log-loss of predictions from a traditional binary model for home-win indicators to those from the proposed binary-normal model (jointly modeling home-win and yards-per-play) and to those from two binary-Poisson models (home-win and sacks, home-win and fumbles) using a sign test. The sign test allows us to measure whether a significant ($\alpha=0.05$ in this section) majority of games experience improved prediction under an alternative joint model. Likewise, a sign test is used to compare the absolute residuals for score, fumble, and sack predictions to measure improvement due to the joint modeling of the binary outcome with these responses. Pragmatically, a significant sign test on the median difference between the log-losses from two models indicates that wagers based on the preferred model would be expected to perform significantly better when equal wagers are placed for all games.

As discussed in section~\ref{sec:computation}, there are some cases in which the Hessian of the model parameters is not positive-definite at convergence. This can indicate instability in the parameter estimates; however, the predictions resulting from these models are still useful, as demonstrated by the improved performance on (hold-out) test data. This is a generalization of the behavior of linear regression models in the presence of multicollinearity.

In this section, we  refer to the (bivariate) normal-binary model as NB. PB0 refers to the Poisson-binary model with no game-level random effect, while PB1 indicates the Poisson-binary model with a game-level random effect. B, N, P0, and P1 refer to the individual binary, normal, Poisson with no game effect, and Poisson with a game effect models, respectively. We report whether there is a significant difference between the home and away mean values from the individual models N, P0, and P1 (the home-field effect is significant in all years for the home-win outcomes in model B). These results are interesting since the multiple membership models account for the quality of the opponents that these values were recorded against. The contrasts between the home and away parameters in the mean vector are tested using the estimated Hessian.

This section does not account for the multiple comparisons that are performed when declaring significance across seasons; however, the p-values are reported in 
the tables. Using the tables,
 it is informative to compare the optimal model identified across seasons. For example, the sacks/home-win model shows improved predictions for sacks over the individual model for sacks in each of the eight seasons, even though only two of those improvements are significant. We would expect to see a preference for the individual model in cross-validation if the jointly modeled response were irrelevant, and a uniform distribution on the resulting p-values, as is the case in Table~\ref{tab:fumbles} for the fumble models.

\subsection{Yards per Play and Outcomes}
When jointly modelling the yards per play and outcome, the joint normal-binary (NB) model provides significantly better predictions for Win/Loss outcomes than individual binary model (B) in  all years (see Table~\ref{tab:ypp}). There is slightly weaker evidence of improvement in the fit of the yards per play in the joint model over the individual normal model: comparing the absolute residuals from each model, there is a significant preference (via the sign test) for the joint model in all but one of the eight years (2006).  In all eight seasons presented, there is a significant game location effect: home teams record more yards per play than visiting teams (p-value for all years $<0.0001$). There is a weak correlation between yards per play recorded by opponents within a game, ranging from 0.05 to 0.15.

\begin{table}[htbp]
  \centering
  \caption{Yards per play (YPP) and binary home-win indicators are modelled both individually (N and B respectively) and jointly (NB). 
  ``Best'' Model for YPP indicates which model, N or NB, provided the best YPP prediction measured by the minimum absolute residual for the majority of games in each year.
  ``Best'' Model for W/L indicates which model produces the best prediction, B or NB, measured by log-loss on the predicted win-probabilities from each model for the majority of games in each year. *Indicates a significant preference over comparison model(s).  }
    \begin{tabular}{lll}\\
    \toprule
                &``Best''     &  ``Best''   \\
			  & Model     & Model \\
    Year  & for YPP & for W/L   \\
    \midrule
    2005  & NB   & NB*   \\
    2006  & N      & NB*   \\
    2007  & NB*  & NB*  \\
    2008  & NB     & NB*  \\
    2009  & NB     & NB*   \\
    2010  & NB* & NB*   \\
    2011  & NB    & NB*    \\
    2012  & NB     & NB*   \\
    2013  & NB*& NB*   \\
    \bottomrule
    \end{tabular}%
  \label{tab:ypp}%
\end{table}%

\subsection{Sacks and Outcomes}
The joint model PB0 (Poison-Binary with no game-level random effects) for sacks and home-win indicators significantly outperforms the individual model B with respect to log-loss for the home-win indicators in each year  (see Table~\ref{tab:sacks}). Likewise, PB0 outperforms the individual sack model P0 in each year (significantly so in two years). While the joint modeling of sacks and outcomes improves the predictions of both responses, the inclusion of a game-level effect in the sack model (PB1) leads to worse predictions in each year (significant in all years for log-loss for the outcomes and in three years for the absolute residuals of the sacks). This indicates that there is no intra-game correlation in the number of sacks recorded by opponents. There was a larger frequency of sacks recorded by the home team in each year (significant in four of the eight years: 2007, 2008, 2009, 2011).

\begin{table}[htbp]
  \centering
  \caption{Sacks and binary home-win indicators are modelled both individually with a Poisson model (P0 and B respectively) and jointly (PB0 or PB1).
  ``Best'' Model for Sacks indicates which model, P0, PB0, or PB1,  provided the best sack prediction measured by the minimum absolute residual for the majority of games in each year.
  ``Best'' Model for W/L indicates which model produces the best prediction, B, PB0, or PB1, measured by log-loss on the predicted win-probabilities from each model for the majority of games in each year. *Indicates a significant preference over comparison model(s).
  } 
      \begin{tabular}{lll}\\
    \toprule
                &``Best''     &  ``Best''   \\
			  & Model     & Model \\
    Year  & for Sacks & for W/L   \\
    \midrule
		 2005  & PB0  & PB0*   \\
    2006  & PB0      & PB0*  \\
    2007  & PB0  & PB0*  \\
    2008  & PB0     & PB0*  \\
    2009  & PB0     & PB0*  \\
    2010  &PB0* & PB0*   \\
    2011  & PB0    & PB0*   \\
    2012  &PB0*    & PB0*   \\
    2013  & PB0  & PB0*   \\
    \bottomrule
    \end{tabular}%
  \label{tab:sacks}%
\end{table}%

\subsection{Fumbles and Outcomes}\label{sec:fumb.out}
Joint modeling of fumbles per game along with the game outcome did not lead to significant differences in log-loss for the outcome predictions in any season, nor did it provide any improvement in the predictive accuracy for the number of fumbles  (see Table~\ref{tab:fumbles}). Furthermore, model P0 outperformed model P1 in every season (significantly so in three seasons), suggesting that there is not a substantial correlation between the number of fumbles recorded by opponents within a game. The home-field effect is not significant in any of the years for P0. This suggests that there is not a tendency for teams to fumble more or less often while traveling.
\begin{table}[htbp]
  \centering
  \caption{Fumbles and binary home-win indicators are modelled both individually with a Poisson model (P0 or P1 and B respectively) and jointly (PB0 or PB1).
  ``Best'' Model for Fumbles indicates which model, P0, P1, PB0, or PB1,  provided the best fumble prediction measured by the minimum absolute residual for the majority of games in each year.
  ``Best'' Model for W/L indicates which model produces the best prediction, B, PB0, or PB1, measured by log-loss on the predicted win-probabilities from each model for the majority of games in each year. *Indicates a significant preference over comparison model(s).
  }
      \begin{tabular}{lll}\\
    \toprule
                &``Best''     &  ``Best''   \\
			  & Model     & Model \\
    Year  & for Fumbles & for W/L   \\
    \midrule
		 2005  & P0  & PB0   \\
    2006  & PB0      & PB0  \\
    2007  & P0  & PB0  \\
    2008  & PB0     & B  \\
    2009  & PB0*     & PB0*  \\
    2010  &PB0 & PB0   \\
    2011  & PB0    & PB0   \\
    2012  &PB0    & B   \\
    2013  & PB0  & PB0   \\
    \bottomrule
    \end{tabular}%
  \label{tab:fumbles}%
\end{table}%

By contrast, a logistic regression on the home-win indicators against the number of home fumbles and the number of away fumbles indicates that these are significant predictors for whether the home team will win. Likewise, the home-win indicators significantly improve predictions for the number of home and away fumbles in a Poisson regression. This provides a good contrast between jointly modeling two responses and including one of the responses as a factor in a model for the other: the former searches for a correlation between latent team effects from each of the responses, while the later considers only relationships between the responses on an observation-by-observation basis. In other words, the joint model considers relationships between higher levels in the hierarchy of the models.

\subsection{Scores and Outcomes}
In each year, model PB1 significantly outperforms the predictions for the home-win indicators from both model B and model NB  (see Table~\ref{tab:score}). This observation comes in spite of the fact that the estimated Hessian was nearly singular in each year, due to the nearly linear relationship between team win-propensities, team offensive (score) ratings, and team defensive (score) ratings.  In this case, the conditional model of \citet{annis} provides a more accurate framework for the data generation process, accounting for the deterministic relationship between the two responses. Nevertheless, this situation highlights the utility of jointly modeling responses in general. Despite the parameter instability (via the inflated standard errors resulting from the near-singular Hessian) of the joint model in the extreme case of modeling scores with the home-win indicator, the home-win predictions still show improvement over those from model B. In fact, in each year the score/home-win model significantly outperforms the yards-per-play/home-win model, which in turn outperforms the sacks/home-win model with respect to log-loss for the home-win predictions.
\begin{table}[htbp]
  \centering
  \caption{Scores and binary home-win indicators are modelled both individually with a Poisson model (P0 or P1 and B respectively) and jointly (PB0 or PB1).
  ``Best'' Model for Scores indicates which model, P0, P1, PB0, or PB1,  provided the best score prediction measured by the minimum absolute residual for the majority of games in each year.
  ``Best'' Model for W/L indicates which model produces the best prediction, B, PB0, or PB1, measured by log-loss on the predicted win-probabilities from each model for the majority of games in each year. *Indicates a significant preference over comparison model(s).
  }
      \begin{tabular}{lll}\\
    \toprule
                &``Best''     &  ``Best''   \\
			  & Model     & Model \\
    Year  & for Scores & for W/L   \\
    \midrule
		 2005  & P1  & PB1*   \\
    2006  & PB1      & PB1*  \\
    2007  & P1  & PB1*  \\
    2008  & P1     & PB1*  \\
    2009  & PB1     & PB1*  \\
    2010  &PB1 & PB1*   \\
    2011  & P1    & PB1* \\
    2012  &P1    & PB1*  \\
    2013  & P1  & PB1* \\
    \bottomrule
    \end{tabular}%
  \label{tab:score}%
\end{table}%

P1 outperforms P0 in every year with respect to absolute residuals for the score predictions, significantly so in four of those years. Furthermore, there are significantly better results from PB1 over PB0 in log-loss for home-win predictions in two of the years. Together, these results suggest that there is an important intra-game correlation between opponent scores. Yet, no significant differences appear with respect to the absolute score residuals appear between models N and P1, or between PB1 and P1. This indicates that the normal and Poisson models for scores perform similarly, and that the game-score predictions are not influenced by the joint modeling of the home-win indicators. This last point is unsurprising since the home-win indicators are a discretized version of a difference of the team scores.

\subsection{Estimated Random Effect Covariance Matrices}
For the 2005, the random effect covariance matrices are presented in table \ref{ta:covmat} (the 2006-2013 season are omitted for brevity). 
 The correlation matrices are printed below the covariance matrices.
 Recall, the columns correspond to the ```offensive'' effect, the ``defensive'' effect, the win-propensity effect, and the game-level effect (score-outcome model only). The correlation matrices are printed below the covariance matrices.
The words offensive and defensive appear in quotes as a reminder that the interpretation of these effects depends on the model structure described in section~\ref{sec:model}. For example, in the sacks-outcomes model, we use the number of sacks recorded by the home team (against the visiting quarterback) as the home response, and likewise define the away response. Thus, a larger offensive effect in the sacks-outcomes model for a given team indicates a larger propensity for that team's \textit{defense} to sack the opposing quarterback.

Additionally, the estimate for the win-propensity variance components from the binary-only model are $(0.43, 0.63, 0.65)$ for three increasingly accurate approximations: first-order Laplace, ``partial'' fully exponential Laplace, and fully exponential Laplace \citep{karlcgs}. 
Notice how the estimates for this component from the fumbles-outcomes models are typically similar to the estimate from the first-order approximation (which was used for all of the joint models). This is not surprising, since no significant differences in the outcome prediction accuracy was noted with the joint modeling of fumbles in section~\ref{sec:fumb.out}. By contrast, the sacks-outcomes, yards/play-outcomes, and scores-outcomes models, which were found to produce progressively more accurate outcome predictions, generate progressively larger estimates for the win-propensity variance component. In the same way that the more-accurate fully exponential Laplace approximation tends to correct for the downward bias observed in variance components for a binary response \citep{breslow95,lin96}, the joint-modeling of a relevant response appears to inflate the variance component estimate.\\

\begin{table}

\caption{Random Effect Covariances and Correlation Matrices for the 2005 Season for the Binary model with each game level response (upper triangle). From left to right in each matrix, the columns correspond to the ``offensive'' effect, the ``defensive'' effect, the win-propensity effect, and the game-level effect (score-outcome model only).} \label{ta:covmat}
\centering

\begin{tabular}{ccc} \\
 \toprule
Game-level response & Covariance & Correlation \\   \midrule  \\
Yards Per Play
&$\begin{bmatrix}{}
  0.55 & 0.22 & 0.58 \\
  & 0.35 & 0.44 \\
& & 0.84 \\
  \end{bmatrix}$
&$\begin{bmatrix}{}
  1.00 & 0.50 & 0.85 \\
  & 1.00 & 0.82 \\
 &  & 1.00 \\
  \end{bmatrix}$ \\\\

Sacks
&
$\begin{bmatrix}{}
  0.07 & 0.03 & 0.17 \\
 & 0.09 & 0.14 \\
  &  & 0.55 \\
  \end{bmatrix}$ &
$\begin{bmatrix}{}
  1.00 & 0.42 & 0.89 \\
  & 1.00 & 0.61 \\
   & & 1.00 \\
  \end{bmatrix}$ \\\\
Fumbles
&
$\begin{bmatrix}{}
    \phantom{-}0.02 &   \phantom{-}0.00 & -0.03 \\
    &   \phantom{-}0.01 & -0.05 \\
  & &   \phantom{-}0.44 \\
  \end{bmatrix}$ &
$\begin{bmatrix}{}
  \phantom{-}1.00 & -0.10 & -0.31 \\
  &   \phantom{-}1.00 & -0.79 \\
   & &   \phantom{-}1.00 \\
  \end{bmatrix}$ \\\\

 Scores
& $\begin{bmatrix}{}
  0.11 & 0.07 & 0.35 & 0.00 \\
   & 0.09 & 0.29 & 0.00 \\
  & & 1.20 & 0.00 \\
 &  &  & 0.07 \\
  \end{bmatrix}$

&$\begin{bmatrix}{}
  1.00 & 0.71 & 0.94 & 0.00 \\
  & 1.00 & 0.90 & 0.00 \\
   &  & 1.00 & 0.00 \\
  &  &  & 1.00 \\
  \end{bmatrix}$ \\
\bottomrule 
\end{tabular}
\end{table}

The (1,2) component of the matrices in the score-outcome model gives the correlation between offensive and defensive team score ratings.
It ranges from 0.77 for American college football data to $-0.3$ for the professional basketball (NBA) data (not shown). We would expect to see a moderate positive correlation in the American college football data: if schools are able to recruit good offensive players and coaches, they will likely also be able to recruit good defensive ones. Interestingly, the offensive and defensive team ratings are negatively correlated for the NBA data. This may reflect the fact that offense and defense are played by the same players in basketball.

\begin{figure}
\caption{Football offensive and defensive yards-per-play ratings from the normal-binary model for the 2012 season. The colors and marker sizes indicate the win propensity rating of each team.}
\label{plot:ncaaf}
\centering
\includegraphics[width=5.5in]{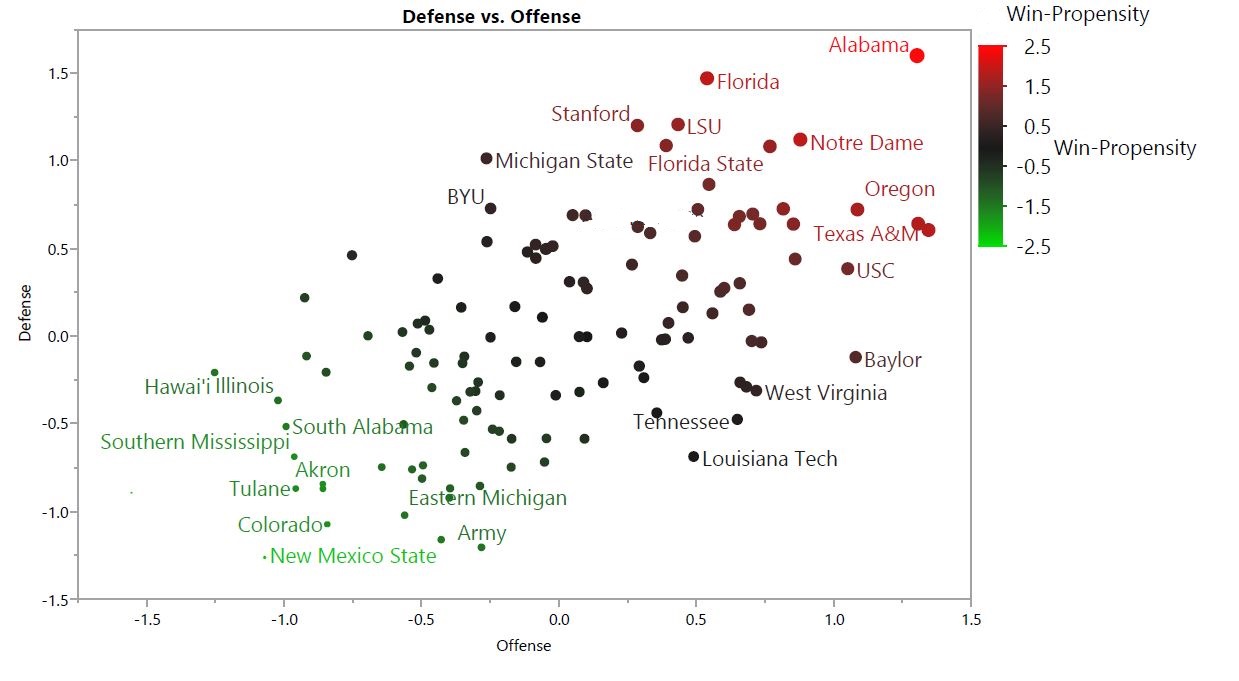}
\end{figure}

Likewise, the (1,2) component of the matrices in the yards/play-outcome model gives the correlation between offensive and defensive team yards-per-play ratings. Figure~\ref{plot:ncaaf} plots the team defensive ratings against the team offensive ratings. The colors and sizes of the team markers correspond to the team win propensity ratings from the normal-binary model. This plot appears similar to the one based on the score-outcome model in Figure~2 of \citet{karlcgs}.

\subsection{Estimated Error Covariance Matrices}
The bivariate normal-binary model provides an estimate of the intra-game correlation between opposing team outcomes. This section uses yards-per-play as the game-level response. The model revealed an intra-game correlation of 0.17 for 2005, 0.04 for 2006, and 0.13 for 2007. 
Hence, there is only weak positive correlation between opposing team yards-per-play within games. There might be a positive relationship due to variance induced by weather conditions, time of day, time in the football season, etc; however, that relationship does not appear to be substantial.

\section{Prediction of NCAA Basketball Tournament Results}\label{sec:ncaapred}
The annual NCAA Division I basketball tournament provides an excellent occasion for sports predictions. The most popular format of tournament forecasting requires a prediction for the winner of each bracket spot prior to the beginning of the first round. By contrast, some contests \citep{contest} require predicted probabilities --  as opposed to discrete win/loss prediction -- of outcomes for each potential pairing of teams. This allows the confidence of predictions to be evaluated while ensuring that a prediction is made for every match that occurs.  We consider the use of the multivariate generalized linear mixed model (\ref{eq:joint}) to produce predicted outcome probabilities that depend on the observed team scores as well as the home-win indicators.

To illustrate the degree to which the model for a response may be influenced by its conditionally independent counterpart in the joint model (\ref{eq:joint}), we jointly model the team scores and (discretized) binary home-win indicators. By jointly modeling the team scores and binary game outcomes, the team win-propensities are influenced by their correlation with team offensive and defensive ratings, thus incorporating information about the scores into the predicted probabilities from the binary sub-model. To demonstrate the benefit of the joint model over the individual binary model, we  compare the fit of the binary and normal-binary models across the most recent 19 tournaments.

\begin{figure}
\caption{Difference in log-loss for the binary (B) and normal-binary (NB) models across years. Dotted lines indicate the 95\% confidence interval for the mean difference.}
\label{plot:logloss2}
\centering
\vspace{.1in}
\includegraphics[trim = 24mm 156mm 73mm 32mm, clip,width=8cm]{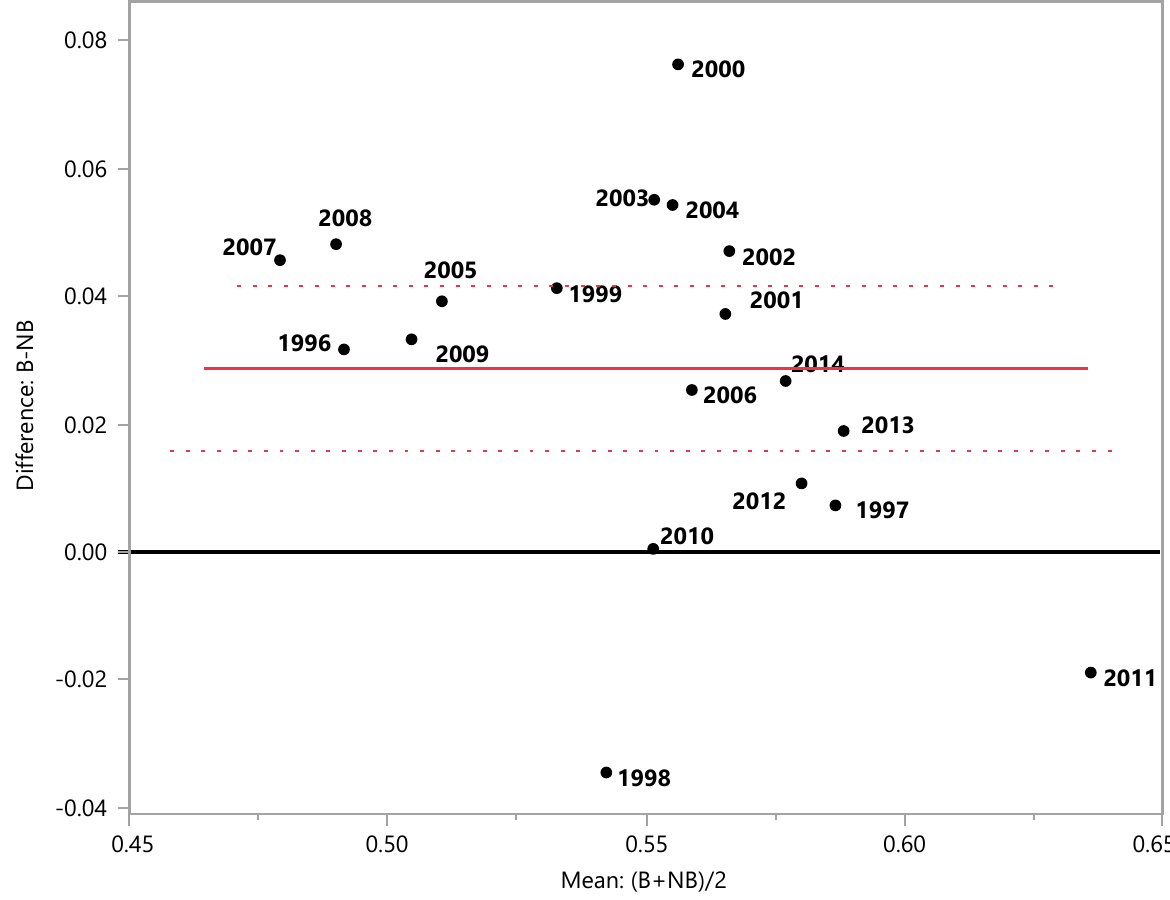}
\end{figure}

Figure~\ref{plot:logloss2} shows that the predicted probabilities produced by joint model NB outperformed (with respect to mean log-loss for tournament games) those produced by model B in all years from 1996-2014 except for two. The p-value for the t-test of the yearly differences in log-loss from the two models is $0.0002$. Thus, the joint model provides a significant improvement in predictive performance for the NCAA tournament by utilizing observed scores while still producing predicted probabilities based on a probit model of outcomes.

\section{Conclusion}
We have developed a combination of multivariate generalized linear mixed models for jointly fitting normal or Poisson responses with binary outcomes. Joint modeling can lead to improved accuracy over separate models for the individual responses.  We have developed and introduced the mvglmmRank package for fitting these models using efficient algorithms.
The mvglmmRank package is not limited to the analysis of football or basketball data: the package is written generally to allow for the analysis of any sport. Differences in scoring patterns within each sport can lead to different patterns of fitted model parameters. For example, basketball produces stronger intra-game score correlations than football. If soccer, baseball, hockey, or other low-scoring sports are to be analyzed, the Poisson-binary model may provide a better fit than the normal-binary model. Furthermore, the estimation routine \citep{karlcgs} is extremely stable, meaning more than two responses could feasibly be modeled jointly.

The process of jointly modeling multiple responses via correlated random effects is useful across a number of applications. For example, \citet{karlcpm} use a similar joint modeling strategy in an analysis of potentially nonignorable dropout, while \citet{broatch10} fit multiple student-level measurements in a joint analysis of a multivariate value-added model.

\appendix
\section{mvglmmRank Package Demonstration}\label{sec:bb}

Scores from the 2012 NCAA Football Bowl Subdivision (FBS) football season were downloaded from \url{github.com/10-01/NCAA-Football-Analytics}. The data include regular season and conference championship outcomes: bowl results are not included.  We chose to ignore inter-division games, and processed the data to match the requirements of mvglmmRank. The processed file is available in the supplementary material. In the following code, the function \verb#file.choose()# is used to select this file from a local directory and load it into the \texttt{game.data} data frame.
\setlength{\topsep}{0pt}

\begin{knitrout}
\definecolor{shadecolor}{rgb}{0.969, 0.969, 0.969}\color{fgcolor}\begin{kframe}
\begin{alltt}
\hlkwd{library}\hlstd{(mvglmmRank)}
\hlcom{#Select football_2012.csv}
\hlstd{game.data} \hlkwb{<-} \hlkwd{read.csv}\hlstd{(}\hlkwc{file} \hlstd{=} \hlkwd{file.choose}\hlstd{())}
\hlstd{game.data}\hlopt{$}\hlstd{home.response} \hlkwb{<-} \hlstd{game.data}\hlopt{$}\hlstd{home.ydsplay}
\hlstd{game.data}\hlopt{$}\hlstd{away.response} \hlkwb{<-} \hlstd{game.data}\hlopt{$}\hlstd{away.ydsplay}
\hlstd{game.data}\hlopt{$}\hlstd{binary.response} \hlkwb{<-} \hlstd{game.data}\hlopt{$}\hlstd{home.win}
\end{alltt}
\end{kframe}
\end{knitrout}

\begin{knitrout}
\definecolor{shadecolor}{rgb}{0.969, 0.969, 0.969}\color{fgcolor}\begin{kframe}
\begin{alltt}
\hlcom{#Output surpressed }
\hlstd{res}\hlkwb{<-}\hlkwd{mvglmmRank}\hlstd{(game.data,}\hlkwc{first.order}\hlstd{=}\hlnum{TRUE}\hlstd{,} \hlkwc{method}\hlstd{=}\hlstr{"NB"}\hlstd{,}
                \hlkwc{Hessian}\hlstd{=}\hlnum{TRUE}\hlstd{)}
\end{alltt}
\end{kframe}
\end{knitrout}

\begin{knitrout}
\definecolor{shadecolor}{rgb}{0.969, 0.969, 0.969}\color{fgcolor}\begin{kframe}
\begin{alltt}
\hlkwd{names}\hlstd{(res)}
\end{alltt}
\begin{Verbatim}[fontsize=\small]
##  [1] "n.ratings.mov"     "n.ratings.offense" "n.ratings.defense"
##  [4] "p.ratings.offense" "p.ratings.defense" "b.ratings"
##  [7] "n.mean"            "p.mean"            "b.mean"
## [10] "G"                 "G.cor"             "R"
## [13] "R.cor"             "home.field"        "actual"
## [16] "pred"              "Hessian"           "parameters"
## [19] "sresid"            "method"
\end{Verbatim}

\begin{alltt}
\hlstd{res}\hlopt{$}\hlstd{parameters}
\end{alltt}
\begin{Verbatim}[fontsize=\small]
##         LocationAway         LocationHome LocationNeutral Site
##            5.4505972            5.8057215            5.5181789
##          Binary mean               R[1,1]               R[2,1]
##            0.2182667            1.4084433            0.1810437
##               R[2,2]               G[1,1]               G[2,1]
##            1.1053667            0.4209778            0.1948942
##               G[3,1]               G[2,2]               G[3,2]
##            0.5965179            0.4346991            0.5927359
##               G[3,3]
##            1.1553399
\end{Verbatim}
\begin{alltt}
\hlstd{res}\hlopt{$}\hlstd{G}
\end{alltt}
\begin{Verbatim}[fontsize=\small]
##        Offense   Defense Win Propensity
## [1,] 0.4209778 0.1948942      0.5965179
## [2,] 0.1948942 0.4346991      0.5927359
## [3,] 0.5965179 0.5927359      1.1553399
\end{Verbatim}
\begin{alltt}
\hlstd{res}\hlopt{$}\hlstd{G.cor}
\end{alltt}
\begin{Verbatim}[fontsize=\small]
##        Offense   Defense Win Propensity
## [1,] 1.0000000 0.4555909      0.8553404
## [2,] 0.4555909 1.0000000      0.8363961
## [3,] 0.8553404 0.8363961      1.0000000
\end{Verbatim}
\begin{alltt}
\hlstd{res}\hlopt{$}\hlstd{R}
\end{alltt}
\begin{Verbatim}[fontsize=\small]
##           Home      Away
## [1,] 1.4084433 0.1810437
## [2,] 0.1810437 1.1053667
\end{Verbatim}
\begin{alltt}
\hlstd{res}\hlopt{$}\hlstd{R.cor}
\end{alltt}
\begin{Verbatim}[fontsize=\small]
##           Home      Away
## [1,] 1.0000000 0.1450977
## [2,] 0.1450977 1.0000000
\end{Verbatim}
\begin{alltt}
\hlkwd{round}\hlstd{(res}\hlopt{$}\hlstd{Hessian[}\hlnum{1}\hlopt{:}\hlnum{4}\hlstd{,}\hlnum{1}\hlopt{:}\hlnum{4}\hlstd{],}\hlnum{2}\hlstd{)}
\end{alltt}
\begin{Verbatim}[fontsize=\footnotesize]
##                      LocationAway LocationHome LocationNeutral Site
## LocationAway               376.12      -249.63               -11.07
## LocationHome              -249.63       345.91                -8.62
## LocationNeutral Site       -11.07        -8.62                25.89
## Binary mean                  4.70        -4.55                 0.07
##                      Binary mean
## LocationAway                4.70
## LocationHome               -4.55
## LocationNeutral Site        0.07
## Binary mean               266.55
\end{Verbatim}
\end{kframe}
\end{knitrout}
\noindent Predictions for future results are available via the \texttt{game.pred} function. To illustrate, we will obtain the predictions for the national championship game, in which Alabama defeated Notre Dame. Alabama averaged 7.25 yards per play, while Notre Dame averaged 5.49 yards per play. The normal-binary model predicted those values to be 5.68 and 4.81, respectively, with a 22.2\% chance of Notre Dame defeating Alabama.

\begin{knitrout}
\definecolor{shadecolor}{rgb}{0.969, 0.969, 0.969}\color{fgcolor}\begin{kframe}
\begin{alltt}
\hlkwd{game.pred}\hlstd{(res,}\hlstr{"Notre Dame"}\hlstd{,}\hlstr{"Alabama"}\hlstd{,}\hlkwc{neutral.site}\hlstd{=}\hlnum{TRUE}\hlstd{)}
\end{alltt}
\begin{Verbatim}[fontsize=\small]
## Normal Distribution for Scores:
## Predicted score for Notre Dame: 4.81
## Predicted score for Alabama: 5.68
##
## Poisson Distribution for Scores:
## N/A for this object.
##
## Binary Distribution for Outcomes:
## Probability of Notre Dame defeating Alabama: 0.222
##
## Normal Distribution for Margin of Victory:
## N/A for this object.
\end{Verbatim}
\end{kframe}
\end{knitrout}

\noindent By contrast, the individual binary (B) model for game outcomes incorrectly predicted a Notre Dame win.

\begin{knitrout}
\definecolor{shadecolor}{rgb}{0.969, 0.969, 0.969}\color{fgcolor}\begin{kframe}
\begin{alltt}
\hlcom{#Output surpressed }
\hlstd{res2} \hlkwb{<-} \hlkwd{mvglmmRank}\hlstd{(game.data,}\hlkwc{first.order}\hlstd{=}\hlnum{TRUE}\hlstd{,} \hlkwc{method}\hlstd{=}\hlstr{"B"}\hlstd{,}
                   \hlkwc{Hessian}\hlstd{=}\hlnum{TRUE}\hlstd{)}
\end{alltt}
\end{kframe}
\end{knitrout}

\begin{knitrout}
\definecolor{shadecolor}{rgb}{0.969, 0.969, 0.969}\color{fgcolor}\begin{kframe}
\begin{alltt}
\hlkwd{game.pred}\hlstd{(res2,}\hlstr{"Notre Dame"}\hlstd{,}\hlstr{"Alabama"}\hlstd{,}\hlnum{TRUE}\hlstd{)}
\end{alltt}
\begin{Verbatim}[fontsize=\small]
## Normal Distribution for Scores:
## N/A for this object.
##
## Poisson Distribution for Scores:
## N/A for this object.
##
## Binary Distribution for Outcomes:
## Probability of Notre Dame defeating Alabama: 0.625
##
## Normal Distribution for Margin of Victory:
## N/A for this object.
\end{Verbatim}
\end{kframe}
\end{knitrout}

\clearpage

\bibliographystyle{DeGruyter}
\bibliography{JBAKbib}

\end{document}